\documentclass[11pt]{article}

\usepackage[margin=1in]{geometry}
\usepackage[authoryear]{natbib}

\usepackage{graphicx}
\usepackage{float}
\usepackage{caption}
\usepackage{subcaption}
\usepackage{color}
\usepackage{hyperref}

\usepackage{authblk}

\usepackage{lipsum}   

\title{Learning from Anatomy: Supervised Anatomical Pretraining (SAP) for Improved Metastatic Bone Disease Segmentation in Whole-Body MRI%
  \thanks{This research was funded by \emph{Fonds Wetenschappelijk Onderzoek} (FWO), grant 1S43623N.}}

\author[1,3,5]{Joris Wuts\thanks{Corresponding author: \texttt{joris.sebastiaan.wuts@vub.be}}}
\author[1,2]{Jakub Ceranka}
\author[3]{Nicolas Michoux}
\author[3]{Frédéric Lecouvet\thanks{Frédéric Lecouvet and Jef Vandemeulebroucke should both be considered last author.}}
\author[1,2,4]{Jef Vandemeulebroucke}

\affil[1]{Vrije Universiteit Brussel, Department of Electronics and Informatics, Pleinlaan 2, 1050 Brussels, Belgium}
\affil[2]{imec, Kapeldreef 75, 3001 Leuven, Belgium}
\affil[3]{Cliniques universitaires Saint Luc \& Institut de Recherche Expérimentale et Clinique (IREC), UCLouvain, Avenue Hippocrate 10, 1200 Brussels, Belgium}
\affil[4]{Universitair Ziekenhuis Brussel, Department of Radiology, Laarbeeklaan 101, 1090 Brussels, Belgium}
\affil[5]{Fonds Wetenschappelijk Onderzoek (FWO), Rue de Louvain 38, 1000 Brussels, Belgium}

\date{}  

\begin{document}
\maketitle
\begin{abstract}
The segmentation of metastatic bone disease (MBD) in whole-body MRI (WB-MRI) is a challenging problem. Due to varying appearances and anatomical locations of lesions, ambiguous boundaries, and severe class imbalance, obtaining reliable segmentations requires large, well-annotated datasets capturing lesion variability. Generating such datasets requires substantial time and expertise, and is prone to error. While self-supervised learning (SSL) can leverage large unlabeled datasets, learned generic representations often fail to capture the nuanced features needed for accurate lesion detection.

In this work, we propose a Supervised Anatomical Pretraining (SAP) method that learns from a limited dataset of anatomical labels. First, an MRI-based skeletal segmentation model is developed and trained on WB-MRI scans from healthy individuals for high-quality skeletal delineation. Then, we compare its downstream efficacy in segmenting MBD on a cohort of 44 patients with metastatic prostate cancer, against both a baseline random initialization and a state-of-the-art SSL method.

SAP significantly outperforms both the baseline and SSL-pretrained models, achieving a normalized surface Dice of 0.76 and a Dice coefficient of 0.64. The method achieved a lesion detection $F_{2}$ score of 0.44, improving on 0.24 (baseline) and 0.31 (SSL). When considering only clinically relevant lesions larger than 1~ml, SAP achieves a detection sensitivity of 100\% in 28 out of 32 patients.

Learning bone morphology from anatomy yields an effective and domain-relevant inductive bias that can be leveraged for the downstream segmentation task of bone lesions. All code and models are made publicly available.
\end{abstract}
\section{Introduction}
Metastatic bone disease (MBD) is a frequent complication in advanced cancers, particularly in prostate and breast malignancies \citep{Padhani2014-jd,clezardin2021bone}. Quantitative biomarkers such as total tumor volume, diffusion volume, and the apparent diffusion coefficient (ADC) have emerged as crucial tools for assessing the severity of the disease and monitoring treatment response, underscoring the importance of early and accurate volumetric assessments of skeletal lesions \citep{VanNieuwenhove2022,oprea-lager2021bone}. However, precise manual segmentation of these often heterogeneous lesions remains a labor-intensive and error-prone process, motivating the development of automated and robust segmentation methods \citep{desouza2022standardised}.

Over the past decade, deep learning has become a popular medical image segmentation technique, yet its success depends on the availability of large, high-quality annotated datasets. In the case of metastatic disease, data collection is particularly challenging due to the sparse distribution of small lesions, the inherent variability in expert annotations, and the diverse appearance of lesions in different anatomical locations throughout the skeleton. This label scarcity problem has motivated the adoption of transfer learning paradigms, where knowledge learned in one domain (e.g., large-scale natural or medical images) is applied to a related task.

Recently, self-supervised learning (SSL) has shown promise by learning general-purpose feature representations from unlabeled data through proxy tasks such as inpainting, rotation prediction, or contrastive instance discrimination \citep{Chen2020-ff,ZHOU2021101840}. Despite these advances, SSL-derived representations tend to be agnostic to the fine-grained anatomical details critical for delineating subtle pathological deviations; moreover, according to an extensive recent study in medical imaging \citep{Zhang2023Dive}, SSL may offer marginal or even negative returns, and the influencing factors are poorly understood. This observation motivates alternative strategies for pertaining networks which can serve as a robust foundation for downstream segmentation tasks. Among these, supervised pretraining on easily annotated healthy anatomy is appealing.

In this work, we propose a novel Supervised Anatomical Pretraining (SAP) strategy and validate its performance on the task of segmentation of bone lesions from whole-body MRI (WB-MRI). Our approach leverages the features of skeletal anatomy by first training a skeletal segmentation model on WB-MRI scans from healthy subjects and subsequently fine-tuning it for lesion segmentation in the skeleton. This strategy is inspired by the clinical observation that radiologists routinely use their understanding of normal anatomy to identify subtle anomalies.

Our contributions are threefold: \begin{itemize}
\item We introduce SAP, a novel supervised anatomical pretraining method designed to explicitly leverage healthy anatomy priors to improve the accuracy of metastatic lesion segmentation in WB-MRI.
\item We validate the effectiveness of SAP through comprehensive comparisons, demonstrating that it significantly outperforms both randomly initialized models and existing state-of-the-art self-supervised methods in lesion detection and segmentation tasks in low data regimes.
\item We extensively evaluate SAP's performance specifically in detecting and segmenting metastatic bone disease in WB-MRI, underscoring its clinical utility and broader potential to enhance diagnostic accuracy and patient monitoring across diverse oncologic imaging scenarios.
\end{itemize}

By explicitly harnessing healthy anatomical features, our method offers a compelling alternative to conventional transfer learning strategies, addressing both the challenges of data scarcity and the limitations of generic SSL representations.

\section{Related Works}

In this section, we review the literature motivating our study. First, we summarize recent advances in supervised deep learning methods for the detection and segmentation of MBD in skeletal MRI. Next, we describe the state-of-the-art in SSL approaches, highlighting their relevance to medical imaging tasks across multiple modalities. Finally, we discuss recent supervised pretraining approaches that leverage large annotated 3D datasets, serving as a foundation for subsequent transfer learning in downstream segmentation tasks.
\paragraph{}
Early work by Ceranka et al. introduced an automated computer-aided diagnosis pipeline tailored for detecting and segmenting metastatic bone disease from WB-MRI. By implementing a robust preprocessing pipeline, this approach surpassed prior methods, attaining an $F_2$-score of 0.50 for detection and a lesion segmentation Dice coefficient of 0.53. This study underscored the importance of rigorous preprocessing to enhance downstream performance, setting a benchmark for automated analysis of MBD in WB-MRI ~\citep{Ceranka2020-jg}.

Later, Kim et al. proposed a method specifically for segmenting bone metastases in spinal MRI using a multicenter dataset ~\citep{Kim2024}. Their evaluation of various combinations of MRI sequences has demonstrated superior performance with a 2D U-Net architecture integrating non-contrast $T_{1}$-weighted and contrast-enhanced fat-suppressed images. Their model achieved a sensitivity of 0.93 per lesion and a Dice coefficient of 0.70 on lesions larger than 1 $cm^3$. These results confirm the potential of deep learning approaches to substantially enhance clinical evaluation of metastatic bone disease, potentially streamlining radiological workflows and improving diagnostic accuracy.

\paragraph{}
Self-supervised learning has emerged as a dominant paradigm for pretraining medical imaging models without requiring extensive manual annotations. Prominent approaches include masked image modeling (e.g., Masked Autoencoder, MAE \citep{He2022MaskedAA}), contrastive learning (e.g. SimCLR \citep{Chen2020-ff}), and generative tasks such as image restoration through inpainting and deformation (e.g., Model Genesis~\citep{ZHOU2021101840}). Recent benchmarks such as the OpenMIND initiative~\citep{wald2024openmind}, systematically compared SSL methods across diverse medical tasks and imaging modalities, demonstrating that fine-grained reconstruction pretext tasks primarily benefit segmentation performance, whereas contrastive instance discrimination approaches typically enhance classification outcomes.~\cite{Zhang2023Dive} provided an extensive empirical evaluation of predictive, contrastive, generative, and hybrid SSL frameworks, highlighting that no single method uniformly excels across all tasks. Their findings stress the importance of aligning SSL pretext tasks closely with downstream objectives, architectural consistency, and domain relevance, particularly emphasizing SSL’s advantage in scenarios characterized by class imbalance.

In line with these insights, ~\cite{Tang2021-rd} introduced a tailored SSL framework specifically for 3D medical image analysis, employing a Swin UNETR architecture that integrates masked volume inpainting, 3D rotation prediction, and contrastive learning. This combination of global context modeling and local representation learning has demonstrated state-of-the-art performance on benchmarks such as the Medical Segmentation Decathlon (MSD)~\citep{Antonelli2022} and Beyond the Cranial Vault (BTCV) datasets~\citep{Gibson2018}. Given the dual requirement of our downstream tasks, both segmentation and detection, we adopt this SSL framework as a benchmark against our method.

\paragraph{}
Complementary to SSL approaches, supervised pretraining leveraging densely annotated datasets has consistently delivered robust performance in medical segmentation tasks. Established examples such TotalSegmentator~\citep{Wasserthal2022-ls} provide extensive, voxel-wise annotations across a wide range of anatomical structures in CT images. Recently, ~\cite{Li2024} further expanded upon these supervised pretraining paradigms through SUPREM, a unified supervised pretraining framework trained on AbdomenAtlas 1.1, comprising more than 9,000 CT volumes annotated with 25 anatomical structures and additional pseudo-labels for various tumor types. Employing both CNN-based architectures (e.g., SegResNet) and transformer-based models (e.g., Swin UNETR), they illustrated how large-scale, fully annotated datasets could significantly enhance downstream segmentation tasks, particularly in scenarios with limited labeled data.

\paragraph{}
In contrast to the generalized pretraining approaches commonly employed in SSL frameworks such as OpenMIND and supervised methods like SUPREM, our proposed method explicitly targets metastatic bone disease segmentation using WB-MRI. Generalized pretraining methods typically rely on large-scale, modality-generic datasets that predominantly encompass common imaging sequences and anatomical domains. Consequently, they often lack adaptability to specialized imaging setups such as WB-MRI, which uniquely combine modalities such as $T_{1}$-weighted and diffusion-weighted imaging (DWI) into multi-channel inputs. Adapting these generic pretrained models directly to specialized imaging modalities is challenging, primarily because datasets of adequate scale and quality to replicate the original pretraining schemes are rarely available. Our approach addresses these limitations by specifically aligning the pretraining strategy to the downstream clinical application, using a relatively small yet modality- and anatomy-specific dataset. Leveraging high-resolution anatomical representations from healthy skeletal anatomy within the same imaging domain as the downstream task allows our model to efficiently capture both global anatomical context and local structural detail essential for the simultaneous detection and segmentation of bone lesions. This targeted strategy reduces the dependence on large-scale generic datasets, enhances modality and domain consistency, and thus potentially leads to improved diagnostic accuracy and transferability.

\section{Methodology}

\subsection{Data Acquisition}

\begin{figure}[ht]
    \centering
    \includegraphics[width=\linewidth]{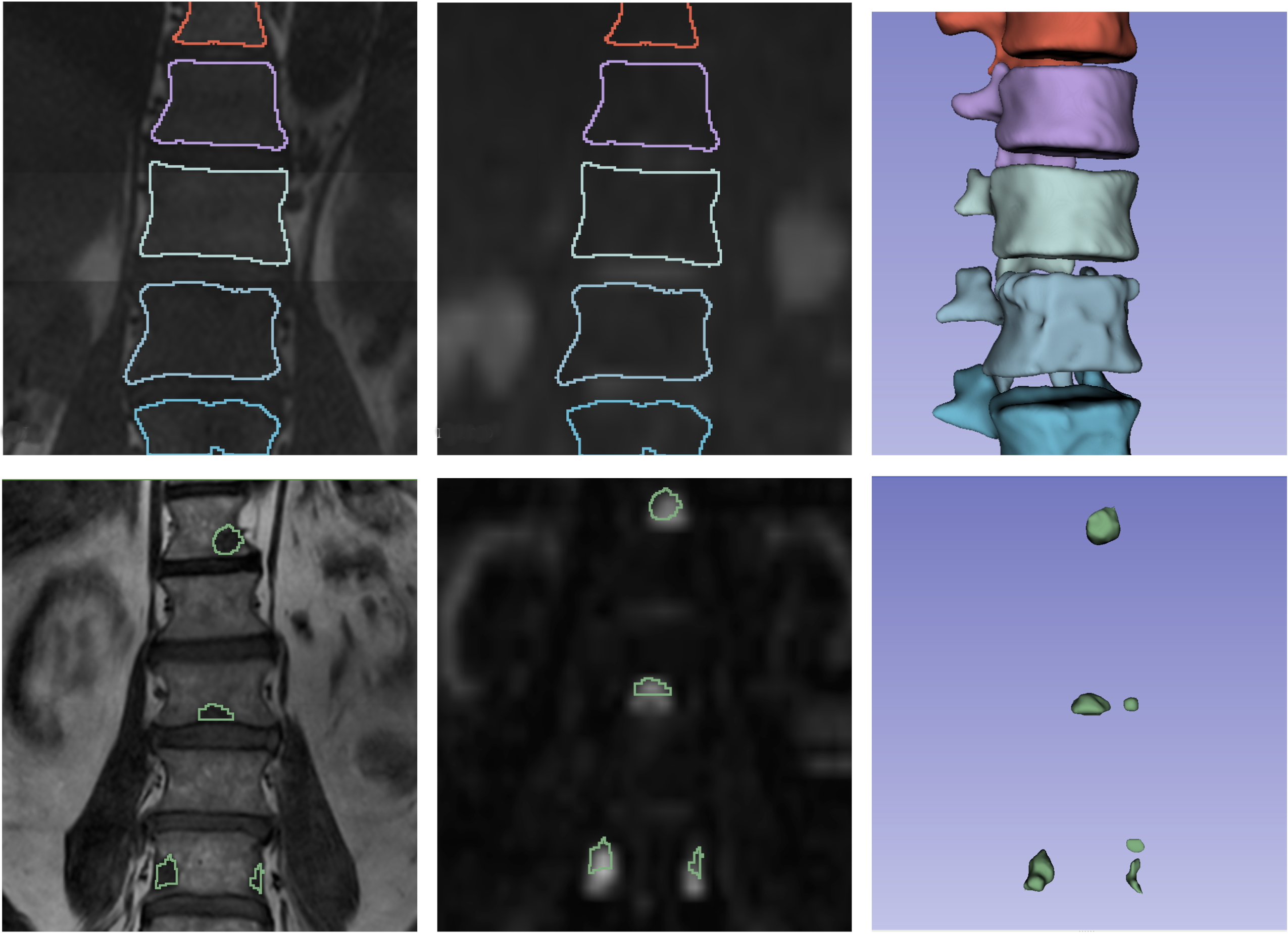}
    \caption{Visualization of three imaging channels of coronal views of the thoracic spine: \textbf{(Left)} $T_{1}$-weighted MRI, \textbf{(Middle)} $b_{1000}$ diffusion-weighted image, and \textbf{(Right)} 3D render of the segmentation mask. The top row represents the thoracic spine of a healthy subject with a multi-label skeletal segmentation, while the bottom row depicts the thoracic and lumbar spine of a patient with metastatic bone lesions. Areas with low $T_{1}$ signal intensity together with  high $b_{1000}$ signal intensity correspond to viable bone metastases.}

    \label{fig:multi_modal}
\end{figure}

We acquired two WB‐MRI datasets: an anatomical dataset for supervised anatomical pretraining, and a pathology dataset to assess the downstream efficacy of the method. The anatomical dataset comprises 24 healthy volunteers \citep{Michoux2021-rk}, yielding 72 multi-parametric scans ($T_{1}$ and DWI) obtained at three different research institutes. Manual skeletal annotations were initially performed on one scan per subject using 3D Slicer \citep{Fedorov2012} by trained interns who underwent training on skeletal representation in anatomical MRI sequences. These initial segmentations were subsequently propagated to the remaining scans via a piecewise linear, bone-specific registration pipeline. Final manual adjustments were then conducted by a radiology expert with seven years of experience in skeletal representations using WB-MRI. The pathological dataset includes WB‐MRI scans from 44 advanced prostate cancer patients with confirmed skeletal metastases, where lesions in the central spine, pelvis, femurs, and clavicles were delineated according to MET‐RADS‐P recommendations \citep{Padhani2017-wq}. Manual delineations were performed in ITK-SNAP \citep{Yushkevich2016-ap} by a medical imaging specialist with seven years of experience in WB-MRI and subsequently validated during a consensus session with an expert radiologist specializing in oncologic imaging and bone metastases identification using multiparametric WB-MRI. Preprocessing, following the pipeline described in \cite{Ceranka2023-kc}, involved noise and bias correction, intensity normalization, rigid registration, and whole-body stitching to form a two-channel ($T_{1}$ and DWI $b_{1000}$) image stack. Figure~\ref{fig:multi_modal} shows the three channels ($T_{1}$, DWI $b_{1000}$, and segmentation mask) for a representative subject from both datasets. Detailed data acquisition and preprocessing protocols are provided in Appendix~\ref{app:data-details}.

\subsection{Experimental setup: metastatic bone disease}

For our supervised training experiments, we employed the Swin UNETR architecture as described by \cite{Hatamizadeh2022-vn}. In our implementation, the network is configured with parameters found in Table \ref{tab:supervised_architecture}. 

\begin{table}[ht]
\centering
\caption{Supervised Training Model Architecture Parameters}
\label{tab:supervised_architecture}
\begin{tabular}{p{0.25\textwidth} p{0.65\textwidth}}
\hline
\textbf{Component} & \textbf{Details} \\ \hline
\textbf{Input} & Patch size: $96 \times 96 \times 96$ voxels; In channels: 2; Out channels: 2 \\[1ex]
\textbf{Encoder} & 
\begin{itemize}
    \item Feature Size: 48
    \item Encoder Depths: [2, 2, 6, 2]
    \item Window Size: $7^3$
    \item Patch Size: $2^3$
    \item Drop Rate: 0.1; Attention Drop Rate: 0.1; Drop Path Rate: 0.1
\end{itemize} \\[1ex]
\textbf{Decoder} & Standard U-Net style decoder using transposed convolutions and skip connections. \\[1ex]
\textbf{Optimizer \& LR} & AdamW with a base learning rate of $1 \times 10^{-5}$ and weight decay of $1 \times 10^{-5}$ \\[1ex]
\textbf{Scheduler} & Cosine annealing learning rate scheduler with a linear warmup phase (warmup epochs = 40, total epochs = 400) \\[1ex]
\textbf{Data Augmentation} &  Random crop, Random flips, intensity shifts.\\ \hline
\end{tabular}
\end{table}

We evaluated three distinct supervised training strategies for metastatic bone disease segmentation, all using the underlying Swin UNETR architecture. A summarization of the three methods is given, highlighting their differences in model initialization and the subsequent fine-tuning procedure on the MBD dataset:

\subsubsection{Random Initialization (Baseline)}
In the baseline approach, the entire network is initialized with random weights and trained directly on the MBD dataset. This method serves as a reference to quantify the benefits of any pretraining strategy.

\subsubsection{Self Supervised Pretraining (SSL)}
For the SSL pretrained strategy, we adopt the default self-supervised learning implementation as described in the original work of \cite{Tang2021-rd}. In this pipeline, the model is first pretrained on unlabeled WB-MRI volumes using a combination of inpainting, rotation prediction, and contrastive learning objectives. After SSL pretraining, the encoder weights are transferred to the MBD segmentation model. The decoder is newly initialized, and the entire network is fine-tuned on the metastatic dataset.

\subsubsection{Supervised Anatomical Pretraining (SAP)}
In our proposed SAP approach, we leverage a previously trained model on healthy skeleton segmentation, which captures explicit bone contours in nine anatomical regions: the cervical, thoracic, and lumbar spine; pelvis; femurs; humeri; scapulae; clavicles; and sternum. For the MBD segmentation task, we transfer both the encoder and decoder weights from this model; however, the final classification head that is tailored for multi-class segmentation is discarded and replaced by a randomly initialized binary segmentation head. During fine-tuning on the metastatic dataset, we apply a differential learning rate strategy: a lower initial learning rate (e.g., $2 \times 10^{-6}$) is used for the encoder to preserve the learned anatomical features, while a higher initial learning rate (e.g., $1 \times 10^{-5}$) is used for the decoder to better adapt to the binary lesion segmentation task. 

\hspace{\linewidth}

\subsection{Experimental setup: Skeletal segmentation}
Using the healthy volunteer dataset (24 subjects, 71 scans) described in Section 3.1, we trained a supervised skeletal segmentation model to delineate nine key anatomical skeletal regions: the cervical, thoracic, and lumbar spine; pelvis; femurs; humerus; scapulae; clavicles; and sternum. The model employs the same Swin UNETR architecture and identical training hyperparameters as used for the metastatic bone disease segmentation network (see Table \ref{tab:supervised_architecture}). In particular, a multi-class segmentation head with ten output channels was used to learn the aforementioned bone classes. A fixed training and validation set was made ensuring the scans originating from the same subject are within the same split. In total, we trained the model on 63 scans and validated it on the remaining 8.  This approach yielded a robust anatomical model that captures healthy skeletal morphology, providing the foundation for our SAP initialization in the downstream lesion segmentation task.

\subsection{Evaluation Metrics, Cross-Validation, and Statistical Testing}

Table~\ref{tab:eval_metrics} summarizes the evaluation metrics used in our study. We distinguish between detection metrics which assess the ability to correctly identify lesions on a per-patient basis, and segmentation metrics which evaluate the quality of the lesion delineation on correctly detected lesions. The evaluation metrics were selected based on the Metrics Reloaded framework \citep{Maier-Hein2024-bv}, which also provides more detailed definitions of all metrics. During evaluation, we report detection metrics both across all lesions and specifically for lesions larger than 1 $ml$, the latter reflecting the subset most relevant for clinical decision-making and patient follow-up in accordence to the Response Evaluation Criteria in Solid Tumors (RECIST, \citep{Eisenhauer2009-kv}).

Additionally, to assess the performance of the skeletal segmentation pretraining, we computed the segmentation Dice coefficient. In this context, we report class-specific Dice scores for each bone region, and also a global binary Dice score by merging all bone regions into a single foreground class (whole-skeleton vs. background), reflecting overall accuracy of skeleton segmentation.

\begin{table}[ht]
\centering
\caption{Overview of evaluation metrics. The selection was guided by the Metrics Reloaded work \cite{Maier-Hein2024-bv}}
\label{tab:eval_metrics}
\begin{tabular}{llp{0.48\textwidth}}
\hline
\textbf{Metric} & \textbf{Type} & \textbf{Description} \\ \hline
FPPI (False Positives per Image) & Detection & Average number of false positive detections per image. \\[1ex]
Lesion-Level $F_2$-Score & Detection & Combination of precision and recall with extra emphasis on recall. \\[1ex]
Sensitivity & Detection & Proportion of true lesions that are correctly detected. \\[1ex]
FROC & Detection & Free-response Receiver Operator curve (computed up to 15 FPPI). \\[1ex]
Dice Similarity Coefficient (DSC) & Segmentation & Quantifies volumetric overlap between predicted and ground-truth segmentations (computed on correctly detected lesions). \\[1ex]
Normalized Surface Dice (NSD) & Segmentation & Measures the agreement of segmentation boundaries with a 2-pixel tolerance to account for annotation uncertainty. \\ \hline
\end{tabular}
\end{table}

A 7-fold cross-validation strategy was employed on the metastatic dataset to ensure robust performance evaluation. In each fold, the dataset was divided patient-wise into 6 folds for training and 1 fold for testing. Additionally, one split was reserved for determining an optimal decision boundary for binarizing the segmentation probability maps for each method. The threshold criteria used was to optimize the $F_2$-score; all metrics are reported on the remaining 6 test splits.

For statistical testing, we first assessed normality of all metric distributions using the Shapiro–Wilk test \citep{Shapiro1965}, which indicated that none were normally distributed. Accordingly, comparisons between methods were performed using non-parametric tests. For patient-level (detection) metrics, where data are naturally paired across methods, the Wilcoxon signed-rank test \citep{Wilcoxon1945} was used. For lesion-level (segmentation) metrics, an unpaired Wilcoxon test (Mann-Whitney U test \citep{Mann1947}) was applied since the set of detected lesions varies between methods. Bonferroni correction \citep{Armstrong2014-pv} was applied to adjust for multiple comparisons, and p-values and the number of samples were reported where applicable.

\subsection{Implementation Details}
The model was implemented using the MONAI framework \citep{Jorge_Cardoso2022-jp} built on PyTorch. Training and inference were conducted on two NVIDIA Ampere GPUs with 80\,GB of memory using torch Distributed Data Parallel for multi-GPU processing. All source code, including pretrained and finetuned models, will be made publicly available to support further research in this area on \href{https://github.com/jwutsetro/SAP}{\url{https://github.com/jwutsetro/SAP}
}.

\section{Results}

\subsection{Healthy Skeleton Segmentation}
\label{sec:results_skeleton}
We evaluated a Swin UNETR model trained with random weight initialization on healthy WB-MRI scans to segment nine distinct bone categories. On a validation set of eight WB-MRI scans, the model achieved a mean Dice similarity coefficient of 0.90 for binary segmentation of the skeleton. Table~\ref{tab:bone_dice_horizontal} reports mean Dice scores for each bone category.

\begin{table}[ht!]
\centering
\caption{Mean Dice scores for 9 distinct bone categories and a full skeleton evaluation.}
\label{tab:bone_dice_horizontal}
\begin{tabular}{cccccccccc}
\hline
\textbf{Cervical} & \textbf{Thoracic} & \textbf{Lumbar} & \textbf{Pelvis} & \textbf{Femur} & \textbf{Humerus} & \textbf{Scapula} & \textbf{Clavicula} & \textbf{Sternum} & \textbf{Full Skeleton} \\ \hline
$0.90$ & $0.94$ & $0.93$ & $0.80$ & $0.90$ & $0.67$ & $0.70$ & $0.83$ & $0.85$ & $0.90$ \\ \hline
\end{tabular}
\end{table}

\begin{figure}[ht]
    \centering
    \includegraphics[width=0.95\linewidth]{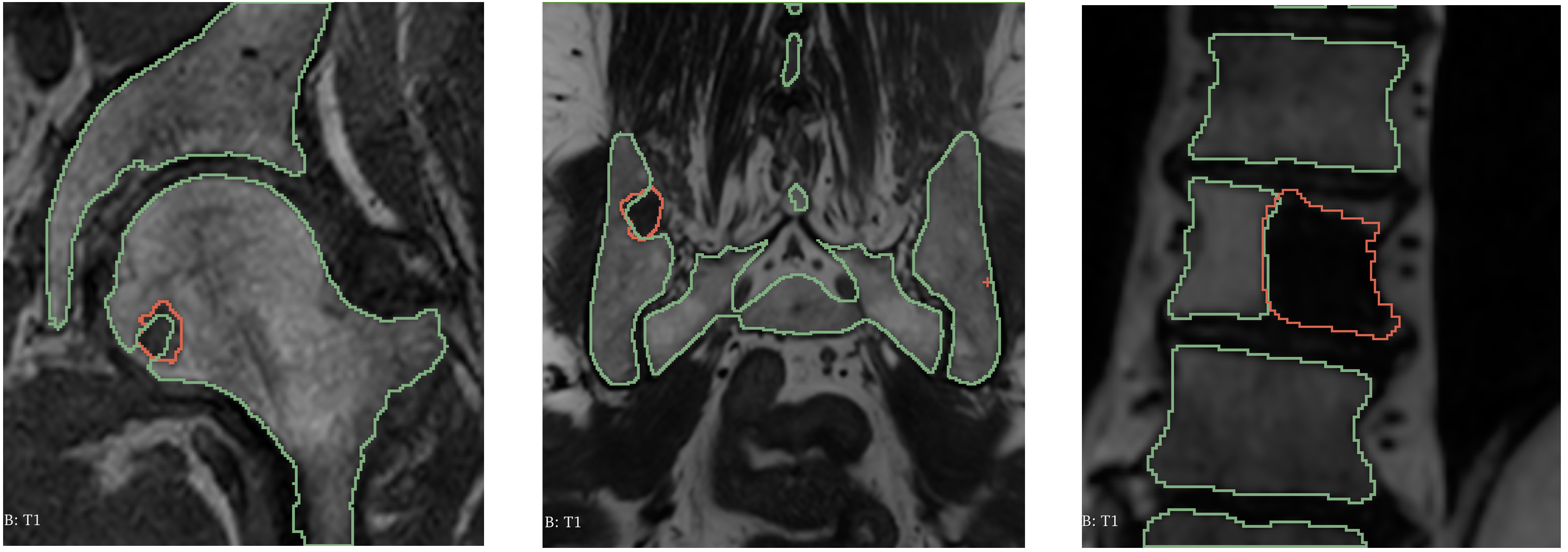}
    \caption{Coronal $T_{1}$ weighted views obtained from three different anatomical locations in three patients with metastatic bone disease showing manual lesion segmentation (red contours) and automated skeletal predictions (green contours). From left to right, the images display a femur with one metastatic lesion, a pelvis with one lesion, and a thoracic vertebra with one lesion. }
    \label{fig:healthy_skeleton_3views}
\end{figure}

Figure~\ref{fig:healthy_skeleton_3views} displays three coronal $T_{1}$-weighted images from patients with metastatic bone disease overlayed with the skeleton prediction maps. These examples demonstrate that while the model robustly delineates most healthy bone structures, its performance deteriorates in regions affected by metastatic infiltration. Specifically, the model systematically fails to include bone tissue affected by metastases in the overall skeletal segmentation, and healthy tissue adjacent to lesions is not accurately delineated.

\subsection{Initial Feature Representation Prior to Finetuning}
\begin{figure}[h]
    \centering
\includegraphics[width=0.65\linewidth]{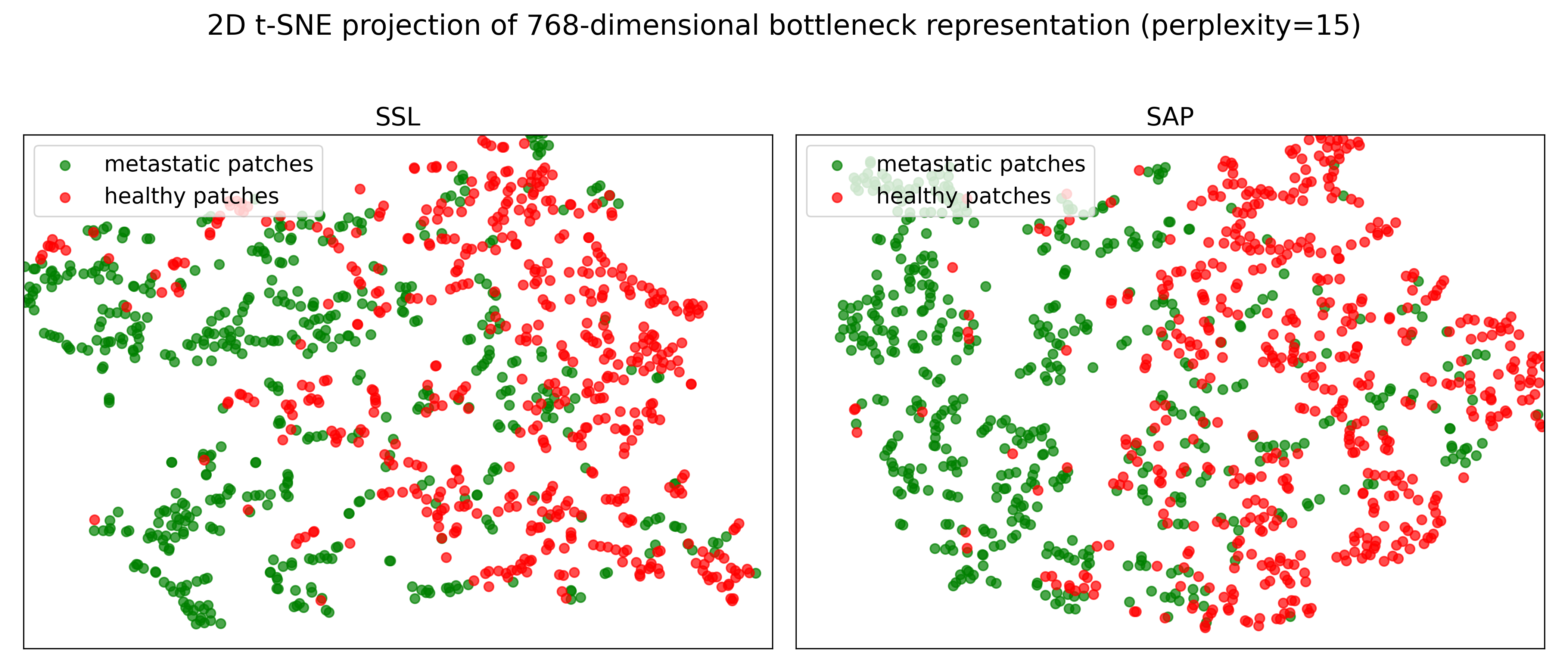}
    \caption{Two-dimensional t-SNE projections of the 768-dimensional bottleneck features (perplexity = 15) learned via two different pretraining strategies prior to fine-tuning on a metastatic bone disease dataset. The left panel shows the distribution of patches learned under SSL, and the right panel shows the distribution from SAP. Green points correspond to patches labeled as positive for metastatic bone disease, whereas red points correspond to negative patches.}
    \label{fig:PCA_plot}
\end{figure}

We visualized the feature representations derived from the SSL and SAP models using a 2D t-SNE projection to capture the global structure of the 768-dimensional bottleneck features of the Swin UNETR. These features were extracted from the models prior to any finetuning on the MBD dataset, thereby reflecting the initialized models' inherent capability to discriminate between lesions and healthy patches. For both methods, all lesion patches and a randomly selected equal number of healthy patches were processed by the models. As shown in Figure~\ref{fig:PCA_plot}, the SAP method exhibits a better class separation compared to the SSL method. The corresponding cluster metrics computed on the full feature space reveal similar inter-cluster (centroid) distances for the two methods (SSL: 9.85, SAP: 9.85). However, the intra-cluster distances differ. For SSL, the average intra-cluster distances are 21.31 for positives and 21.37 for negatives, whereas for SAP they are 10.02 for positives and 10.44 for negatives. Both models are thus positioning the centroids of the positive and negative clusters at comparable distances, however the features from the SAP model are substantially more compact. Together, the PCA pair plot and cluster metrics illustrate that before finetuning, the initialized SAP model is more effective at discriminating positives from negatives compared to the SSL model. 
  
\subsection{Metastatic Bone Disease Segmentation}
\label{sec:results_mbd}
For the MBD segmentation task, we compared three training methods as described in section 3.2. All models were fine-tuned on 44 prostate cancer patients with metastatic lesions and evaluated via 7-fold cross-validation.

\subsubsection{Quantitative Evaluation Metrics}
\begin{figure}[h!]
    \centering
    \includegraphics[width=\linewidth]{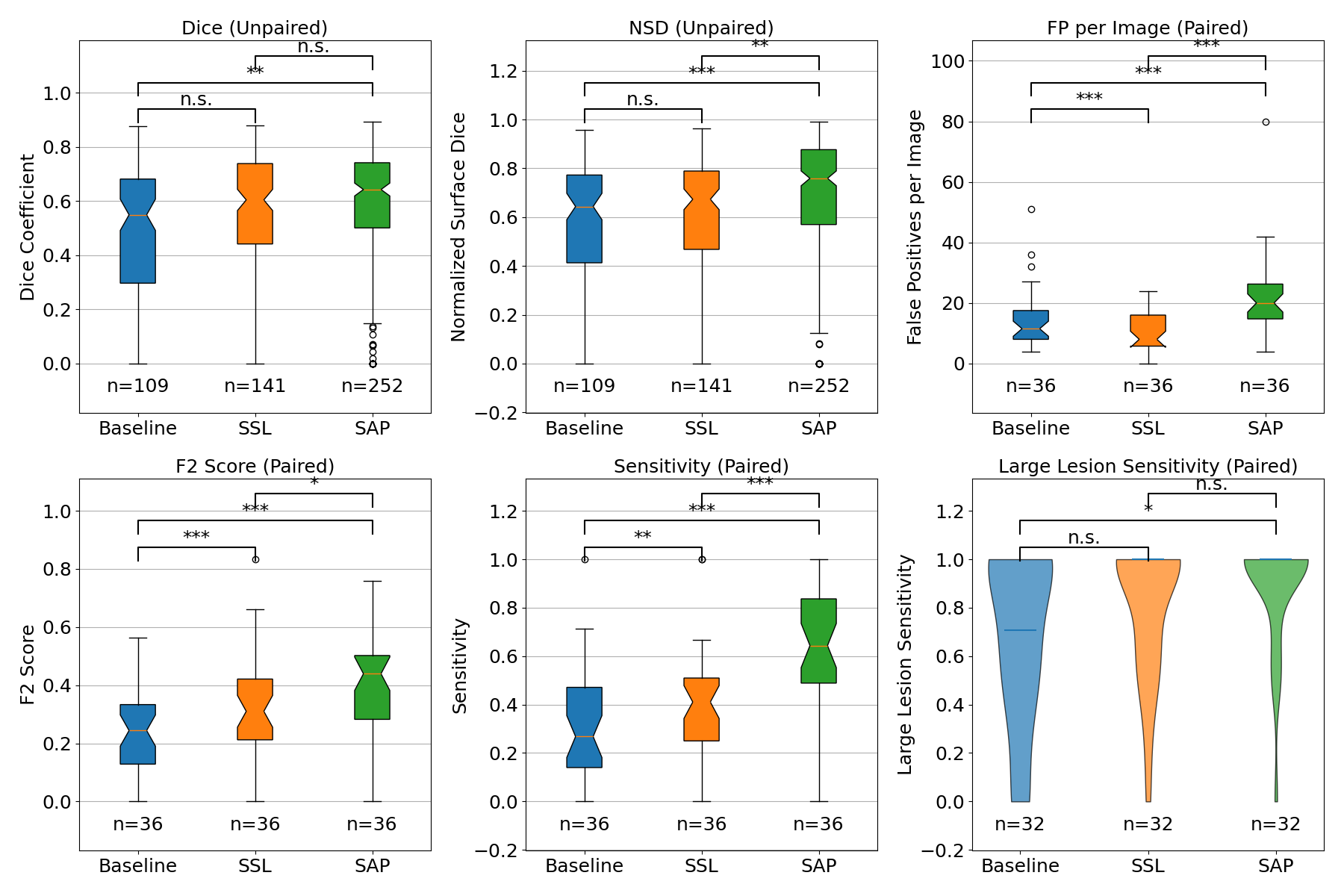}
    \caption{Comparison of detection and segmentation metrics for MBD across three methods (Baseline, SSL, and SAP). Metrics include Dice Coefficient, normalized surface Dice (NSD), false positives per image, $F_2$-Score, detection sensitivity, and large lesion detection sensitivity. Statistical significance is denoted by * (p<0.05), ** (p<0.01), and *** (p<0.001), with 'n.s.' indicating no significance. For the segmentation metrics, the number of detected lesions per method that where included in the distribution is indicated with n. For the detection metrics, n indicates the number of patients included in the distribution.}
    \label{fig:boxplots}
\end{figure}

Figure~\ref{fig:boxplots} presents box plots summarizing both detection and segmentation metrics for the baseline, state-of-the-art and proposed method. The obtained results reveal a consecutive progression in detection and segmentation performance: the SSL model improves upon the baseline, and the SAP approach further improves upon the SSL method.  Worth noting is that the SAP model consistently outperforms both the baseline and SSL-pretrained models in key segmentation metrics (e.g., Dice, Normalized Surface Dice) and detection sensitivity.

The median sensitivity increases from 0.27 for the baseline to 0.41 for SSL, and reaching 0.64 with SAP, while the $F_2$-score improves from 0.244 (baseline) to 0.31 (SSL) and 0.44 (SAP). Consistent increase in quantitative validation metrics were also observed in segmentation metrics, with Dice scores increasing from 0.55 (baseline) to 0.60 (SSL) and 0.643 (SAP), and the Normalized Surface Dice (NSD) improving from 0.64 to 0.67 and 0.76, respectively. Although the SAP model exhibits a higher false positive rate (20.0 FP per image) compared to the baseline (11.5) and SSL (8.0), this increase in FPPI comes with an improved sensitivity and segmentation quality. See Table \ref{tab:all_metrics} for a comprehensive overview of the exact median values and interquartile ranges.

\begin{table}[ht!]
\centering
\caption{Overview of median detection and segmentation metrics for Baseline, SSL, and SAP. The 25 and 75\% percentiles are shown in parentheses.}
\label{tab:all_metrics}
\begin{tabular}{lccc}
\hline
 & \textbf{Baseline} & \textbf{SSL} & \textbf{SAP} \\
\hline
\multicolumn{4}{c}{\textbf{Detection Metrics}} \\
\hline
\multicolumn{4}{c}{\emph{All Lesions}} \\
\textbf{Sensitivity}        & 0.27 (0.14–0.47)     & 0.41 (0.25–0.51)     & 0.64 (0.49–0.84)     \\
\textbf{FP per Image}       & 11.50 (8.00–17.50)   & 8.00 (5.75–16.00)    & 20.00 (14.75–26.25)  \\
\textbf{$F_2$ Score}        & 0.24 (0.13–0.33)     & 0.31 (0.21–0.42)     & 0.44 (0.28–0.50)     \\
\hline
\multicolumn{4}{c}{\emph{Lesions >1 ml}} \\
\textbf{Sensitivity}        & 0.71 (0.50–1.00)     & 1.00 (0.67–1.00)     & 1.00 (1.00–1.00)     \\
\textbf{FP per Image}       & 0.00 (0.00–1.00)     & 0.00 (0.00–0.00)     & 0.00 (0.00–1.00)     \\
\hline\hline
\multicolumn{4}{c}{\textbf{Segmentation Metrics}} \\
\hline
\textbf{Dice}               & 0.55 (0.30–0.68)     & 0.60 (0.44–0.74)     & 0.64 (0.50–0.74)     \\
\textbf{NSD}                & 0.64 (0.41–0.77)     & 0.67 (0.47–0.79)     & 0.76 (0.57–0.88)     \\
\hline
\end{tabular}
\end{table}

\subsubsection{FROC Analysis}
\begin{figure}[ht!]
    \centering
    \includegraphics[width=\linewidth]{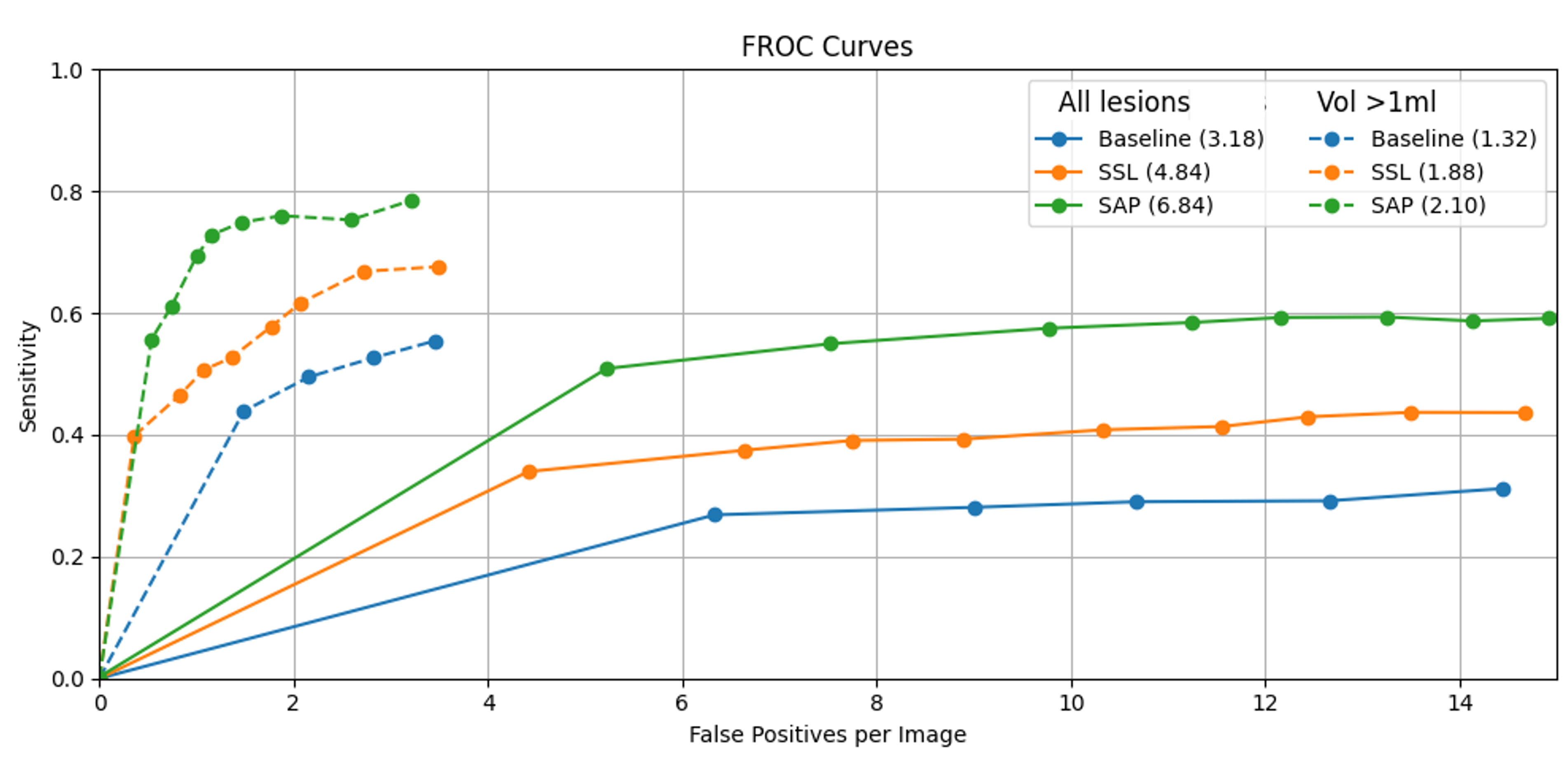}
    \caption{Combined FROC curves for lesion detection: solid lines show all lesions up to 15 FPPI, and dashed lines show lesions with volume > 1 $ml$ up to 4 FPPI, for the Baseline, SSL, and SAP methods. The FROC‐AUC is numerically integrated under each piecewise‐linear curve to its respective FPPI limit and reported in the legend.}

    \label{fig:froc_curves}
\end{figure}

Figure~\ref{fig:froc_curves} illustrates the FROC curves for the three methods, plotting lesion detection sensitivity against false positives per image (up to 15 FPPI). A stepwise improvement is observed: the Baseline method performs worse than the SSL approach, which in turn is outperformed by the SAP model. This trend is confirmed by the higher FROC AUC values (3.18, 4.84 and 6.84 respectivly), with SAP achieving the greatest area under the curve. Moreover, the curves remain nearly parallel, demonstrating that for any given FPPI threshold, sensitivity increases progressively from Baseline to SSL and from SSL to SAP. 
We observe similar trends when restricting the analysis to clinically relevant lesions larger than 1 $ml$ (dashed lines). All three methods achieve higher sensitivities and fewer false positives per image in this setting, reflecting the improved detectability of larger lesions. The relative ranking of the methods remains unchanged, with SAP consistently outperforming both SSL and Baseline approaches across the full range of FPPI values.

\subsubsection{Sensitivity and Lesion Volume Analysis}
\begin{figure}[htbp]
    \centering
    \includegraphics[width=\linewidth]{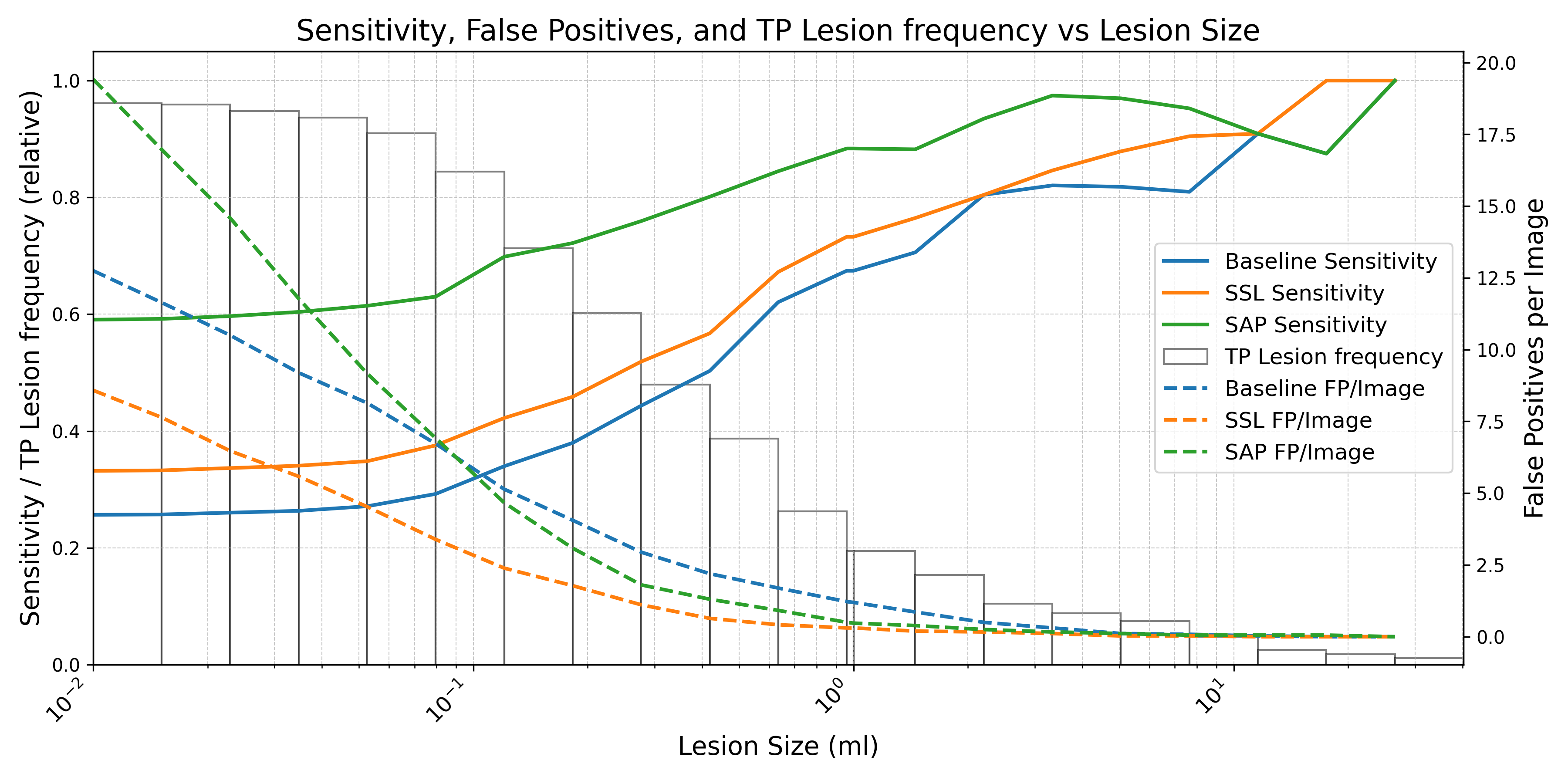}
    \caption{Combined evaluation of sensitivity and false positives per image (FP/image) for three methods (Baseline, SSL, SAP) as a function of lesion size. Sensitivity (solid lines) and FP/image (dashed lines) are plotted against lesion volume, with the left y-axis for sensitivity and the right y-axis for FP/image. Sensitivity and FP/image are cumulative metrics, considering all lesions larger than a given volume. The white histogram represents the cumulative distribution of true lesion volumes providing insight into lesion volume prevalence.}
    \label{fig:sensitivity_vs_volume}
\end{figure}

To evaluate the influence of lesion size on detection performance, we analyzed sensitivity, FPPI, and the frequency distribution of lesion volumes (Figure~\ref{fig:sensitivity_vs_volume}). All methods exhibit reduced sensitivity for small lesions. However, when considering only lesions larger than 1 $ml$ (clinically relevant), detection sensitivity shows a clear stepwise improvement—from 0.68 for the Baseline, to 0.72 for SSL, and up to 0.87 for SAP. The volume histogram reveals these clinically relevant lesions account for only 19\% of all annotations. In contrast, most false positives are associated with small lesions; when restricting predictions to lesions above 1 $ml$, mean FPPI values drop to 1.19, 0.31, and 0.47 for Baseline, SSL, and SAP, respectively. For smaller volume thresholds, FPPI increases more rapidly for SAP, resulting in the highest overall FPPI. These findings underscore the trade-off between sensitivity and FPPI and emphasize the importance of selecting an appropriate minimum prediction volume.

\subsubsection{Qualitative visualisation.}
\begin{figure}[htbp]
    \centering
    \includegraphics[width=\linewidth]{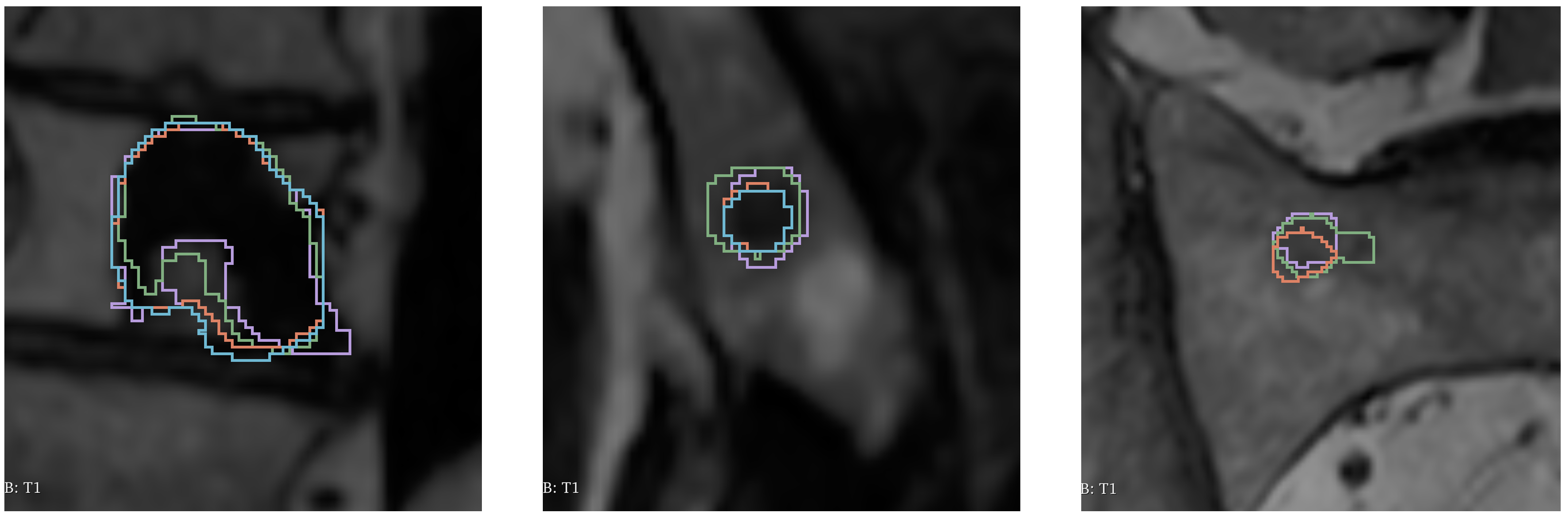}
    \caption{Coronal $T_{1}$‐weighted MR images from subjects with metastatic lesions are shown with manual annotations in purple. Overlaid are the predicted segmentations from three training paradigms: Baseline (blue), SSL (orange), and SAP (green). The panels (left to right) correspond to lesions in the cervical spine, right pelvis, and sacrum, which are representative of the 25th, 50th, and 75th percentiles of Dice scores (based on the SAP model), respectively, illustrating the variability in segmentation performance across methods.}
    \label{fig:output_predictions}
\end{figure}

Figure \ref{fig:output_predictions} illustrates a qualitative comparison of segmentation results produced by three training paradigms: Baseline, Self-Supervised Learning, and the proposed Supervised Anatomical Pretraining Projection. Three lesions, chosen to represent the 25th, 50th, and 75th percentiles of Dice scores (as computed with the SAP model), are shown on coronal $T_{1}$‐weighted MR images with manual annotations outlined in purple.

In the left panel, a lesion in the thoracic spine is detected by all methods. The segmentation quality progressively improves from Baseline to SSL and finally to SAP, although all models tend to oversegment the inferior concave portion of the lesion. The middle panel shows a lesion in the right pelvis (coxal bone); here, the SAP model slightly oversegments the lesion, whereas the Baseline and SSL models tend to undersegment it. In the right panel, a lesion in the sacrum is not detected by the Baseline model, while the SSL and SAP methods capture the lesion with relatively large undersegmentation and oversegmentation, respectively, compared to the ground truth lesion volume.

\subsubsection{Results on Metastatic Bone Disease Segmentation} Our experimental results indicate that the Baseline method is not able to capture the complex morphology of skeletal lesions, resulting in lower segmentation accuracy and detection performance. The SSL approach, offers moderate improvements by leveraging unlabeled data, but does not incorporate the anatomical priors necessary for precise lesion delineation. The SAP strategy significantly outperforms the other methods by achieving higher segmentation accuracy and improved lesion detection metrics.

\section{Discussion}

Our study demonstrates that the proposed supervised anatomical pretraining method significantly outperforms both random initialization and self-supervised learning. The experiments were performed on a challenging segmentation task for metastatic bone disease on multi-modal WB-MRI. However, this approach can potentially be easily adapted to other segmentation tasks. In this section, we discuss the implications of these findings, their alignment with prior work, and potential avenues for further research.

\subsection{Comparison with Existing Work}
\label{sec:discussion_comparison}
\subsubsection{Skeletal Segmentation}
Our skeletal segmentation model demonstrates robust performance across diverse anatomical regions, comparable to recent approaches such as the works of ~\cite{Wennmann2022-xs}, ~\cite{Ceranka2019}, and ~\cite{Huang2024-segmentanybone}. Wennmann et al. reported binary Dice scores of 0.91–0.94 using WB-MRI at lower out-of-plane resolution (5–8 mm), while Ceranka et al. employed a multi-atlas segmentation approach achieving an average Dice of $0.887\pm0.011$ for selected bones. SegmentAnyBone, a recent universal bone segmentation model, achieved an average Dice of 0.86 over multiple bones and MR sequences.

Our method differentiates itself by utilizing higher-resolution WB-MRI scans (1.2 mm out-of-plane resolution), allowing precise delineation of intricate anatomical structures such as posterior vertebral elements. Additionally, our evaluation includes anatomies omitted or less accurately captured in prior studies, such as the scapula, clavicle, and sternum, which inherently exhibit lower Dice scores due to their small size and high surface-to-volume ratio, a limitation documented~\citep{Reinke2021-lj}.
Overall, our model’s enhanced resolution and extensive anatomical coverage support effective transfer to downstream tasks such as the segmentation of metastatic bone lesions, typically characterized by small lesion volumes.

\subsubsection{Supervised Anatomical Pretraining vs. Self-Supervised Learning}
Recent studies have employed transfer learning to overcome the challenge of limited pathology-specific labeled data. In particular, self-supervised learning has attracted considerable attention in medical imaging for its ability to leverage large volumes of unlabeled data to learn robust feature representations~\citep{Tang2021-rd,ZHOU2021101840}. However, our findings indicate that while SSL methods excel at capturing general image features, they may not sufficiently acquire the domain-specific structural knowledge essential for precise bone lesion delineation. This is evidenced by our results presented in section \ref{sec:results_mbd}. 

Moreover, our analysis of the initial feature representations, extracted prior to any finetuning, revealed that the SAP model inherently produces more compact and discriminative features for lesions and healthy skeleton. We observed that although both models exhibit similar inter-cluster distances (SSL: 9.847, SAP: 9.851), the intra-cluster variability is markedly lower for SAP compared to SSL (21.31 for positives and 21.37 for negatives). This suggests that even at initialization, the SAP model possesses a stronger inductive bias for distinguishing lesions from healthy anatomy.

These observations are consistent with trends reported in the SUPREM framework proposed by \cite{Li2024}, which demonstrated that supervised pretraining on an equivalent dataset size outperforms SSL pretraining. In our work, the SSL approach required 101 GPU hours for training, whereas SAP achieved superior performance with only 5.5 GPU hours of pretraining. Although a direct comparison is challenging, given that SUPREM employs a CT-based framework that does not directly translate to WB-MRI, our approach distinguishes itself by focusing on the explicit learning of healthy anatomical representations. This focus provides a more targeted inductive bias, effectively enabling the model to differentiate subtle pathological deviations in downstream segmentation tasks.

\subsubsection{Metastatic Bone Disease Segmentation and Detection}
Our approach yields substantial improvements in the detection and segmentation of metastatic bone disease, particularly for larger, clinically relevant lesions. Notably, our supervised anatomical pretraining strategy achieved  a 100 \% sensitivity for large lesions on 28 out of 32 test patients, with an average of only 0,47 false positives per image. While our method enhances the detection of smaller lesions as well, with an overall sensitivity of 0.644, reliably identifying these lesions remains a challenge.

When comparing our results to related work, our findings remain competitive despite differences in anatomical coverage. For example, \cite{Kim2024} reported a Dice coefficient of 0.70 and a per-lesion sensitivity of 0.83 using a U-Net-based approach on spinal MRI. In contrast, our method yields a segmentation Dice of 0.64 and a sensitivity of 0.64 when considering predictions without a minimum volume threshold. However, it is important to note that our dataset includes lesions not only in the spine but also in the pelvis, femur, and clavicles, thereby increasing the anatomical diversity and complexity compared to studies focused solely on spinal metastases. Including these additional regions is necessary for an exhaustive skeletal screening in daily oncology practice Furthermore, our lesion size distribution indicates that we have incorporated a greater number of small, difficult-to-detect lesions. When focusing exclusively on large lesions (greater than 1 $ml$), our reported sensitivity is 0.87 compared to 0.94 in the work of \cite{Kim2024}, showing the robustness of our method for clinically relevant lesions despite the challenges posed by smaller lesions.

Additionally, many of the false positives generated by the SAP model are related to systematic anatomical deviations—such as fractures, scars, and metabolically inactive lesions from previous treatment cycles. While technically false positives, these errors are often clinically acceptable because they reflect interpretable anatomical variations rather than random inaccuracies. Similarly, occasional failures of the healthy skeletal segmentation in regions near lesions (see Figure~\ref{fig:healthy_skeleton_3views})) align with known clinical anomalies. This highlights an important advantage of SAP: its understandable inductive bias. Specifically, SAP first learns to segment anatomical structures, and then, based on this anatomical understanding, segments pathological features such as metastasis. Intermediate results illustrate how anatomical segmentation can fail near pathological regions, offering insights into the model's internal behavior. This interpretability is a  benefit of SAP compared to classical self-supervised learning methods, which rely on less transparent proxy tasks -such as masked auto-encoding  or contrastive learning- and thus function more like black boxes.

\subsection{Limitations and Future Directions}
\label{sec:discussion_limitations}
Despite the promising results, several limitations deserve mentioning. First, our healthy dataset consisted of only 23 volunteers and the MBD dataset included 44 patients, all with advanced prostate cancer. Expanding to larger and more diverse cohorts—with varied scanner types, imaging protocols, and other primary cancers and hematological malignancies—would improve the generalizability of our findings. Future work should explore external validation across institutions and different populations (e.g., multiple myeloma, breast cancer metastases).

\begin{figure}[htbp]
    \centering
    \includegraphics[width=\linewidth]{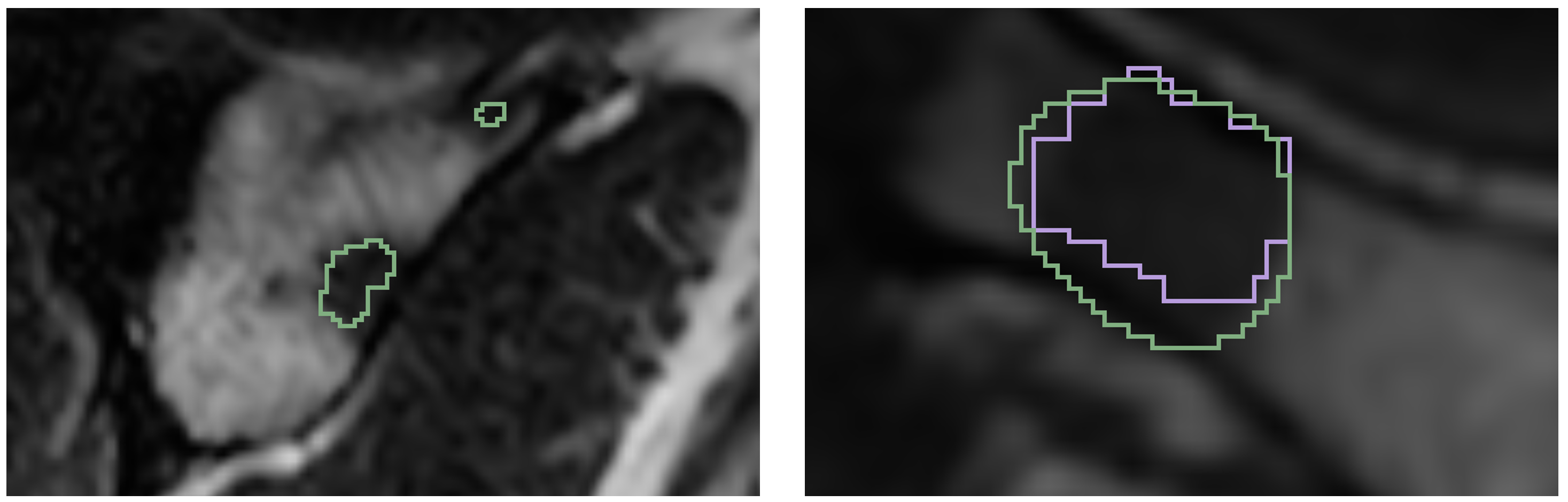}
    \caption{Examples of limitations in the ground truth (GT) lesion annotations. \textbf{Left:} Lesions in the right shoulder fall outside the predefined field of view, leading to their exclusion from the GT and thus causing a true positive lesion to be counted as a false positive. \textbf{Right:} A comparison between the manual lesion boundary (purple) and the anatomical model prediction (green) reveals that the model produces a smoother, more consistent delineation that more accurately captures the hypointense regions on $T_{1}$‐weighted images used to differentiate lesions from surrounding healthy tissue. }

    \label{fig:gt_limitations}
\end{figure}

A second limitation is the quality of the ground truth (GT) lesion annotations. As detailed in Figure~\ref{fig:gt_limitations}, lesions outside the original field of view, that where thus not included in the ground truth segmentation, negatively impact the reported metrics. Additionally inconsistencies were observed between manual and model-predicted lesion boundaries where the latter is sometimes actually outperforming . This inherent variability in manual annotations suggests that further (semi-automated) refinement of the GT may be beneficial.

Finally, many metastases in our dataset were notably small, with a median volume of only \(0.5~\mathrm{ml}\). Although anatomical pretraining significantly improved detection and segmentation, enhancing sensitivity for these low-volume lesions may require specialized strategies, such as the incorporation of custom loss functions or methods that effectively capture features at multiple scales.

\section{Conclusion}
We propose a novel Supervised Anatomical Pretraining (SAP) approach that leverages healthy skeletal anatomy as a robust foundation for metastatic bone disease segmentation in whole‐body MRI. By integrating explicit anatomical priors, SAP significantly enhances lesion detection and segmentation accuracy, outperforming both randomly initialized models and state‐of‐the‐art self‐supervised methods when pretraining dataset sizes are limited. The improvements in sensitivity and segmentation, especially for clinically relevant lesions, show its potential to refine diagnostic workflows and enhance patient monitoring for patients with metastatic bone disease.

Although coverage of the whole spine, pelvis, humerus, femurs, and scapulae already represents a large proportion of the skeleton typically affected by metastases and myeloma (with only the skull and ribs notably absent), further validation on larger, more diverse cohorts is required to establish broader generalizability. Future work should broaden the anatomical scope to include lesions throughout the entire body. Additionally, evaluating SAP’s applicability across different imaging modalities, pathologies, and benchmark datasets could further augment its clinical utility. Overall, our study lays the groundwork for more reliable and interpretable segmentation systems in oncologic imaging, offering a compelling alternative to conventional transfer learning strategies that often function as black boxes.

To enhance transparency and reproducibility, all models, pretrained weights, and source code have been shared publically on https://github.com/jwutsetro/SAP.

\paragraph{\textbf{Acknowledgments}} The resources and services used in this work were provided by the VSC (Flemish Supercomputer Center), funded by the Research Foundation - Flanders (FWO) and the Flemish Government. This study was funded by FWO (grant number 1S43623N).

\paragraph{\textbf{Disclosure of Interests.}}
The authors have no competing interests to declare that are relevant to the content of this article.

\paragraph{\textbf{Data availability}}
The data that has been used is confidential.

\appendix
\section{Detailed Data Acquisition and Preprocessing Protocols}
\label{app:data-details}

This appendix provides a description of the acquisition protocols and preprocessing pipelines for the two whole‐body multi-parametric MRI datasets used in this study. In what follows, we detail both the healthy dataset from the Platform for Imaging in Clinical Research in
Brussels study (PICRIB) and the pathological dataset from advanced prostate cancer patients.

\subsection{Healthy Dataset: The PICRIB Study and Skeletal Segmentation}
Originally, the PICRIB study \citep{Michoux2021-rk} was designed to assess the repeatability and reproducibility of ADC measurements in a multicenter whole‐body MRI protocol. In this work, this standardized dataset is repurposed to support high-quality skeletal segmentation and anatomical pretraining. The healthy cohort consisted of 24 asymptomatic volunteers (10 women and 14 men, aged 23–57 years) recruited from three academic centers in Belgium: Cliniques Universitaires Saint Luc (Université catholique de Louvain), UZ Brussel (Vrije Universiteit Brussel), and Hôpital Erasme, Brussels (Université libre de Bruxelles). Volunteers underwent two scans at one institute and an additional scan at another institute to evaluate both intra- and inter-center reproducibility. All examinations were performed using the same 3.0-T MRI scanner model (e.g., Philips Ingenia 3.0 T,).

\subsubsection*{Manual Skeleton Labeling}
For each subject in the healthy dataset, manual skeletal labeling was performed on one  3D $T_{1}$-weighted scan using 3D Slicer \citep{Fedorov2012}. Initial annotations were generated by trained operators following standardized guidelines, with approximately one subject’s labeling completed per workday. These annotations were subsequently refined by a medical imaging expert with over 7 years of experience in bone anatomy to ensure precise delineation of cortical boundaries.

\subsubsection*{Semi-Automated Piecewise Bone Registration Pipeline}
\begin{figure}[htbp]
    \centering
    \includegraphics[width=\linewidth]{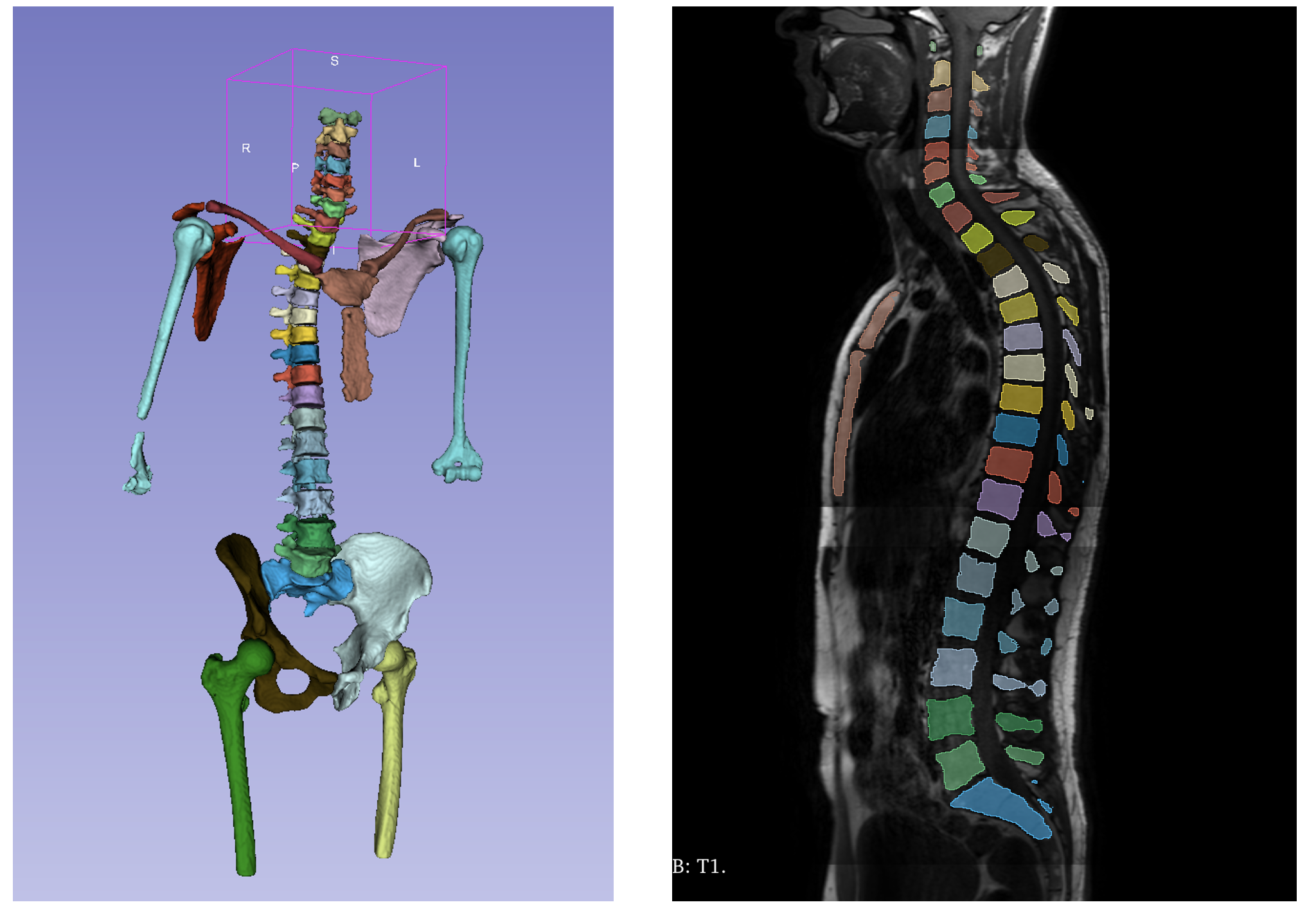}
    \caption{Examples of refined multi-class skeletal segmentation.  \textbf{Left:} 3D render of the included skeletal bones.  \textbf{Right:} A sagittal $T_{1}$ weighted view on the spinal column from a healthy subject with annotated skeleton. }

    \label{fig:Manual_skel}
\end{figure}

To propagate the skeletal annotations across multiple scans per subject, a semi-automated piecewise bone registration pipeline was implemented. The registration protocol consisted of 3 steps:
\begin{enumerate}
    \item \textbf{Global Rigid Registration:} Each subject’s annotated scan was first globally aligned to the target scan to achieve a rough overall correspondence.
    \item \textbf{Localized Bone-Specific Registration:} For each bone, a 10-voxel dilated mask was generated to isolate the region of interest. A localized rigid registration was then applied to precisely align the masked regions.
    \item \textbf{Sequential Spinal Registration:} Special attention was given to the spinal column. Registration was performed sequentially, starting at the sacrum, with each vertebra registered in turn. The transformation from the preceding vertebra was used as the initialization for the next, ensuring robust alignment even for the smaller vertebral bodies.
\end{enumerate}
The proposed piecewise strategy effectively accommodated inter-scanner variability in pixel intensities and subtle anatomical deformations arising from differences in patient positioning and surrounding soft tissue. An example of the refined skeletal annotations can be seen in figure \ref{fig:Manual_skel} 

\subsubsection*{Registration Parameters}
The registration steps were implemented using SimpleElastix \citep{Marstal2016} with the following parameters: the multi resolution registration framework was employed with an Euler-transform and a B-Spline interpolator for registration. The optimizer was adaptive stochastic gradient descent, configured with a maximum of 2000 iterations and 3 resolution levels. Advanced Mattes Mutual Information similarly metric was used. Additional settings included a 10-voxel dilation of the bone masks, an image pyramid schedule of (4,4,2,2,2,1,1,1,1), with 2048 spatial samples per iteration.

\subsection{Pathological Dataset: Multi-Parametric WB-MRI in Advanced Prostate Cancer}
The pathological dataset consists of WB‐MRI scans acquired from 44 advanced prostate cancer patients with confirmed skeletal metastases during routine clinical examinations at Cliniques Universitaires Saint-Luc, Brussels, Belgium. The imaging protocol combined anatomical and functional sequences, namely:
\begin{itemize}
    \item \textbf{Anatomical Sequences:} Acquisitions using either a 3D $T_{1}$-weighted or an in-phase DIXON sequence \citep{Pasoglou2015}.
    \item \textbf{Diffusion-Weighted Imaging (DWI):} Images were acquired at multiple b-values (0, 50, 150, and 1000 s/mm\(^2\)), with the high b-value ($b_{1000}$) images being critical for visual lesion detection in this study \citep{Takahara2004}.
\end{itemize}

\subsubsection*{Manual Lesion Segmentation}
Metastatic lesions in the whole spine, pelvis, femurs, and clavicles were manually delineated on high-resolution $T_{1}$-weighted images following MET‐RADS‐P criteria \citep{Padhani2017-wq}. The manual segmentation was performed using ITK-SNAP and/or 3D Slicer \citep{Yushkevich2016-ap,Fedorov2012} as the primary software tool. Initial annotations were carried out by trained researchers and subsequently refined by a medical imaging specialist with over 7 years of experience in oncology imaging. A total of 403 lesions larger than 50 voxels were annotated with this approach. 

\subsubsection*{Acquisition Parameters}
Table~\ref{tab:imaging_parameters} summarizes the key acquisition parameters for the three primary imaging sequences used in the pathological dataset.

\begin{table}[ht]
\centering
\caption{Acquisition parameters for the anatomical sequences in the pathology dataset.}
\label{tab:imaging_parameters}
\begin{tabular}{lccc}
\hline
\textbf{Parameter}              & \textbf{3DTSE $T_{1}$}          & \textbf{3DGRE $T_{1}$ DIXON}          & \textbf{$b_{1000}$ DWI} \\
\hline
TE (ms)                         & 8                     & 1.15                  & 66 \\
TR (ms)                         & 382                   & 3.6                   & 8421 \\
Matrix Size                     & 480$\times$480        & 432$\times$432        & 192$\times$192 \\
Pixel Spacing (mm)              & 0.65                  & 1.04                  & 2.3 \\
Slice Thickness (mm)            & 1.1                   & 1.5                   & 6.1 \\
\hline
\end{tabular}
\end{table}

\subsection{Comprehensive Preprocessing Pipeline for Whole-Body MRI}
Both datasets were processed using a robust preprocessing framework designed to address the inherent challenges of whole-body MRI, such as spatial distortions and intensity inhomogeneities \citep{Ceranka2023-kc}. The pipeline comprises the following key steps:

\subsubsection*{1. Noise Suppression and Bias Field Correction}
Anisotropic diffusion filtering \citep{Perona1994} was applied to reduce noise while preserving critical edge information, followed by the N4ITK
non-parametric non-uniform intensity normalization
algorithm to correct for low-frequency bias fields \citep{Tustison2010N4ITK}.

\subsubsection*{2. Inter-Station Intensity Standardization}
A linear intensity matching was performed in the overlapping regions of adjacent image stations (Inter-Station Intensity Standardization, as described in the work of \cite{Ceranka2023-kc}) to ensure consistent intensity profiles across the different stations.

\subsubsection*{3. Spatial Registration of DWI Stations}
Rigid registration was employed to align sequential diffusion-weighted imaging (DWI) stations using the pelvis as the reference, thereby correcting for inter-station misalignments \cite{Ceranka2018-sn}.

\subsubsection*{4. Whole-Body Image Reconstruction and Resampling}
The registered stations were stitched together into a continuous whole-body volume via linear interpolation in overlapping areas, ensuring smooth transitions. Subsequently, all DWI images were resampled to match the resolution of the anatomical sequences.

\subsubsection*{5. Inter-Modality Image Registration}
Following the reconstruction of whole-body volumes, DWI images were registered to the anatomical (TSE or GRE $T_{1}$) images using a rigid registration approach supplemented by deformable adjustments as necessary. This step guarantees accurate spatial correspondence between modalities.

\subsubsection*{6. Inter-Patient Intensity Standardization}
A piecewise linear scaling algorithm was applied to standardize intensities across patients. Intensity histograms were aligned based on predetermined percentiles (0, 20, 40, 60, 80, and 95) to mitigate inter-subject variability \citep{Ceranka2023Comparison}.

The integrated preprocessing pipeline has been demonstrated to significantly enhance the performance of our computer-aided diagnosis (CAD) system in prior work of \cite{Ceranka2023-kc}.

\bibliographystyle{cas-model2-names}
\bibliography{reference}

\end{document}